\documentclass[pre,aps,floats,floatfix,superscriptaddress,nofootinbib,showpacs]{revtex4}

\usepackage[english]{babel}

\usepackage{amsmath,amssymb,amsfonts}

\usepackage[dvips]{graphicx}

\usepackage{amsmath}

\usepackage{fancyhdr}

\usepackage{bm}

\newcommand{\be}{\begin{equation}}

\newcommand{\ee}{\end{equation}}

\begin{document}

\title{Energy fluctuations in a randomly driven granular fluid}

\author{Mar\'ia Isabel Garc\'ia de Soria}
\affiliation{Universit\'e Paris-Sud, LPTMS, UMR 8626, Orsay Cedex, F-91405 and
CNRS, Orsay, F-91405}
\author{Pablo Maynar}
\affiliation{Laboratoire de Physique Th\'eorique (CNRS
  UMR 8627), B\^atiment 210, Universit\'e Paris-Sud, 91405 Orsay cedex,
  France}
\affiliation{F\'{\i}sica Te\'{o}rica, Universidad de Sevilla,
Apartado de Correos 1065, E-41080, Sevilla, Spain}
\author{Emmanuel Trizac}
\affiliation{Universit\'e Paris-Sud, LPTMS, UMR 8626, Orsay Cedex, F-91405 and
CNRS, Orsay, F-91405}

\date{\today }

\begin{abstract}

We study the behavior of the energy fluctuations in the stationary state of 
a uniformly heated granular gas. The equation for the one-time two-particle 
correlation function is derived and the hydrodynamic eigenvalues are
identified. Explicit predictions are subsequently worked out
for energy fluctuations. The results explain
Monte Carlo numerical data reported in previous studies
(P. Visco {\it et al}, European Physical Journal B {\bf 51}, 377 (2006)).

\end{abstract}


\maketitle

\section{Introduction}

Recent years have witnessed ongoing interest for the
microscopic and macroscopic properties of granular media. 
In such systems, a simple ingredient --energy dissipation resulting
from collisions-- has far reaching consequences \cite{bte05}, with rich 
phenomenology: 
non-Gaussian velocity distributions \cite{gs95,vne98}, 
non-equipartition of energy \cite{wp02,fm02,bt02,sd06}, or spontaneous
symmetry breaking \cite{sn96,bmgr02,bt03} to name but a few. 
Theoretically, one of the tools 
used to understand this body of phenomena is kinetic theory, which is
extended naturally to these systems by introducing an ``inelastic collision
rule'' in which the energy is not conserved. Although most of
the work carried out until now has focused on the one-particle properties, and
on the study of the corresponding Boltzmann equation, 
it has been shown that correlations are also
important and, as a matter of fact, necessary to
understand the behavior of the system when vortices or cluster are developed
\cite{nebo97,bmr98}, or even in simpler situations where spatial
homogeneity is enforced \cite{bgmr05,bdgm06}.

As a consequence of the dissipation in collisions, the total energy of an
isolated granular system decays monotonically in time. Under certain
conditions, the system reaches a homogeneous cooling state in which
the time dependence of the one-particle distribution function 
is entirely embodied in the kinetic (so-called granular)
temperature, which evolves with time as $t^{-2}$ \cite{gs95,h83}. 
It is experimentally difficult to probe such a regime 
(see however \cite{pdb03,mima08}). 
Nevertheless, it is possible to maintain a
granular system in the fast-flow regime by injecting energy in such a way that
a stationary state is reached. In these states, the energy injected by the
thermostat is compensated by the energy dissipated in collisions. Several 
mechanisms can be introduced in order to get a stationary state. If, for
example, energy is injected by a moving boundary such as a vibrating piston,
the system reaches an
inhomogeneous stationary state \cite{brm2000}. In this work, we will focus on a
granular gas  which is heated uniformly, coupling the velocity of each 
particle to a white noise, this is the so-called  ``stochastic thermostat'' 
\cite{vne98,wm96,plmv99,vnetp99,ptvne01,ms00,mss01,gm02,pvb05,etb06}. 
For this kind of forcing, which is relevant for some 
two-dimensional experimental configurations with a rough vibrating 
piston \cite{peu02},
the system reaches a homogeneous
stationary state after a transient regime. The advantage of 
such a driving mechanism is that it lends itself to theoretical
progress. In this context, the single
particle distribution function has been characterised \cite{vne98} and 
long range correlations predicted from a hydrodynamic treatment
\cite{vnetp99}. More recently,
the fluctuations of the total energy have been analyzed
\cite{vpbvwt06} (see also \cite{affm} for a related numerical study 
in an inhomogeneous system). In Ref. \cite{vpbvwt06}, the second moment of the
total energy fluctuations was 
evaluated by neglecting the correlations, which, by and large,
could not explain the simulation results. The objective of the 
present work is to clarify and quantify the influence
of the inelasticity induced correlations on the total energy fluctuations.
The methods used bear some similarities with those reported in 
\cite{bgmr04}, where it was shown that for the unforced system,
the contribution coming from the 
correlations is of the same order as  that coming from the one-particle
distribution function itself. 




The paper is organized as follows. 
In section \ref{Heated granular gas}, the equation for
the two-particle distribution function is derived, taking
due account of the thermostat while in section \ref{The stationary state},
the results are particularized
to the homogeneous stationary state, that will play the role
of our reference state in subsequent analysis. There we
also summarize the main results already known and 
pertaining to the one-particle distribution
function. In section \ref{Hydrodynamic equations}, we analyze the hydrodynamic
equations for a homogeneous linear perturbation of the reference state,
and obtain the corresponding modes and
eigenvalues, which are finally 
used in section \ref{Energy fluctuations} to meet our objective and 
obtain an explicit expression for the variance of the total energy.

\section{Heated granular gas : two body kinetic description}
\label{Heated granular gas}

We consider a gas of $N$ hard disks (dimension $d=2$) or spheres ($d=3$) 
of mass $m$ and diameter $\sigma$ that collide inelastically with a coefficient
of normal restitution $\alpha$ \cite{bte05}. 
The system is heated uniformly by adding a
random component to the velocity of each particle at equal times
\cite{vnetp99,ptvne01}. The driving 
is implemented in such a way that the time between random kicks is small compared to 
the mean free time. Then, between collisions, the velocities of the particles
undergo a large number of kicks 
due to the thermostat. In addition, we will assume that the
``jump moments'' of the velocities of the particles verify 
\begin{eqnarray}\label{noise}
B_{ij,\beta\gamma}\equiv\lim_{\Delta t\to 0}
\frac{\langle\Delta v_{i,\beta}\Delta v_{j,\gamma}\rangle}{\Delta t}
=\xi_0^2\delta_{ij}\delta_{\beta\gamma}
+\frac{\xi_0^2}{N}(\delta_{ij}-1)\delta_{\beta\gamma},\\
i,j=1,\dots,N \qquad \text{and} \qquad \beta,\gamma=1,\dots,d\nonumber
\end{eqnarray}
where we have introduced 
$\Delta v_{i,\beta}\equiv v_{i,\beta}(t+\Delta t)- v_{i,\beta}(t)$,  
$ v_{i,\beta}(t)$ being the $\beta$ component of the velocity of particle $i$ at
time $t$. We have also introduced the strength of the noise, $\xi_0^2$, and 
$\langle\dots\rangle$, which denotes average over different realizations of the
noise. The non-diagonal terms (corresponding to $i\ne j$ and
$\beta=\gamma$) are necessary in order to conserve the total momentum.

In the dilute limit, assuming molecular chaos, i.e. that no correlations 
exist \emph{between colliding particles}, and that the sizes of the jumps 
due to the thermostat are small compared to the velocity scale on which the 
distribution varies, the equation for the single particle distribution function 
in our system is the Boltzmann-Fokker-Planck equation \cite{vne98,ptvne01}
\begin{equation}\label{ec:boltzmann}
\frac{\partial}{\partial t}f(x_1,t)+L^{(0)}(x_1) f(x_1,t)=J[f\vert f]
+\frac{\xi_0^2}{2}\left(\frac{\partial}{\partial\mathbf{v}_1}\right)^2f(x_1,t),
\end{equation}
where $x_i$ is a short-hand for position-momenta 
coordinates $\left\{\mathbf{r}_i,\mathbf{v}_i\right\}$ and
\begin{equation}
L^{(0)}(x_1)=\mathbf{v}_1\cdot\frac{\partial}{\partial\mathbf{r}_1}.
\end{equation}
The inelastic collision operator $J[f\vert f]$ reads 
\begin{equation}
J[f\vert f]=\sigma^{d-1}\int d\mathbf{v}_2\bar{T}_0(\mathbf{v}_1,\mathbf{v}_2)
f_1(\mathbf{r},\mathbf{v}_1,t)f_1(\mathbf{r},\mathbf{v}_2,t),
\end{equation}
where
\begin{equation}
\bar{T}_0(\mathbf{v}_1,\mathbf{v}_2)=\int 
d\boldsymbol{\hat{\sigma}}\Theta(\boldsymbol{\hat{\sigma}}\cdot\mathbf{g})
(\boldsymbol{\hat{\sigma}}\cdot\mathbf{g})\left[\alpha^{-2}
b_{\boldsymbol{\hat{\sigma}}}^{-1}-1\right],
\end{equation}
with $\mathbf{g}=\mathbf{v}_1-\mathbf{v}_2$ the relative velocity, 
$\Theta$ the Heaviside step 
function, $\boldsymbol{\hat{\sigma}}$ a unit vector joining the centers of 
the particles at contact and $b_{\boldsymbol{\hat{\sigma}}}^{-1}$ an operator 
replacing the velocities $\mathbf{v}_1$ and $\mathbf{v}_2$ appearing on its 
right by the precollisional values 
\begin{eqnarray}
\mathbf{v}_1^*\equiv b_{\boldsymbol{\hat{\sigma}}}^{-1}\mathbf{v}_1 
=\mathbf{v}_1-\frac{1+\alpha}{2\alpha}
(\mathbf{g}\cdot\boldsymbol{\hat{\sigma}})\boldsymbol{\hat{\sigma}},\\
\mathbf{v}_2^*\equiv b_{\boldsymbol{\hat{\sigma}}}^{-1}\mathbf{v}_2
=\mathbf{v}_2+\frac{1+\alpha}{2\alpha}
(\mathbf{g}\cdot\boldsymbol{\hat{\sigma}})\boldsymbol{\hat{\sigma}}.
\end{eqnarray}
The term 
$\frac{\xi_0^2}{2}\left(\frac{\partial}{\partial\mathbf{v}_1}\right)^2f(x_1,t)$
is a diffusive Fokker-Plank term, and is a signature of the external noise. 


As we shall study fluctuations, it is convenient to introduce the 
two-particle distribution function, $f_2(x_1,x_2,t)$. The quantity 
$f_2(x_1,x_2,t)dx_1dx_2$ is defined as the 
number of pairs of particles in which one lies inside the 
differential volume $dx_1$ centred in $x_1$ and likewise, with $dx_2,x_2$ for the
second particle. This definition is easily 
generalized to higher $n$-particle distribution functions, 
$f_n(x_1,\dots,x_n)$. The evolution equation for $f_2$ 
is \cite{ec81,bgmr04}
\begin{eqnarray}\label{ec_f2}
\left[\frac{\partial}{\partial t}+L^{(0)}(x_1)+L^{(0)}(x_2)\right]
f_2(x_1,x_2,t)
=\delta(\mathbf{r}_{12})\sigma^{d-1}\bar{T}_0(\mathbf{v}_1,\mathbf{v}_2)f_2(x_1,x_2,t)
\nonumber\\
+\sigma^{d-1}\int dx_3\left[\delta(\mathbf{r}_{13})\bar{T}_0(\mathbf{v}_1,\mathbf{v}_3)+
\delta(\mathbf{r}_{23})\bar{T}_0(\mathbf{v}_2,\mathbf{v}_3)\right]
f_3(x_1,x_2,x_3,t)+F_{TH}, 
\end{eqnarray}
where we have introduced $F_{TH}$ that accounts for the external driving. 
The evolution equation (\ref{ec_f2})
contains essentially three parts: the free streaming in the left-hand side,
the two terms in the right hand side corresponding to collisions, and the last
term, $F_{TH}$ due to the thermostat. The collisional contribution is split
in one part corresponding to collisions of particles with velocities
$\mathbf{v}_1$ and $\mathbf{v}_2$, and the other which involves collisions of
particles with velocities  $\mathbf{v}_1$ or $\mathbf{v}_2$ with a third
particle with arbitrary velocity, $\mathbf{v}_3$. The collisional
contribution is identical to the one that appears in the absence of forcing 
\cite{bgmr04}. We concentrate now on the new term, 
$F_{TH}$. Assuming that the sizes of the jumps due to the 
thermostat are small compared to the scale in which the distribution $f_2$ 
varies, we can expand $F_{TH}$ in the spirit of the Fokker-Planck 
description \cite{vk92}
\begin{eqnarray}
F_{TH}&\simeq&\frac{1}{2}\sum_{\beta,\gamma=1}^d\sum_{i,j=1}^2
B_{ij,\beta\gamma}
\frac{\partial}{\partial v_{i,\beta}}\frac{\partial}{\partial v_{j,\gamma}}
f_2(x_1,x_2,t)\nonumber\\
&=&\frac{1}{2}\xi_0^2\left[\frac{\partial^2}{\partial v_1^2}+
\frac{\partial^2}{\partial v_2^2}
-\frac{2}{N}\frac{\partial}{\partial\mathbf{v}_1}\cdot
\frac{\partial}{\partial\mathbf{v}_2}\right]f_2(x_1,x_2,t), 
\label{eq:bfp}
\end{eqnarray}
where we have taken into account equation (\ref{noise}), and we have 
explicitly assumed that the jump moments 
$B_{ij,\beta\gamma}$
do not depend on the magnitude of the velocities of the particles.

Let us introduce the two-particle and three-particle correlation functions
through the usual cluster expansion
\begin{equation}
f_2(x_1,x_2,t)=f_1(x_1,t)f_1(x_2,t)+g_2(x_1,x_2,t), 
\end{equation}
and
\begin{eqnarray}
f_3(x_1,x_2,x_3,t)=f_1(x_1,t)f_1(x_2,t)f_1(x_3,t) 
+g_2(x_1,x_2,t)f_1(x_3,t)\nonumber\\
+g_2(x_1,x_3,t)f_1(x_2,t)
+g_2(x_2,x_3,t)f_1(x_1,t)+g_3(x_1,x_2,x_3,t).  
\end{eqnarray}
The equation for the correlation function $g_2(x_1,x_2,t)$ can be obtained 
following the same lines as in references \cite{bgmr04,mgsbt08}.
Neglecting the three-body correlations, $g_3$, in Eq. (\ref{ec_f2}), we
obtain 
\begin{eqnarray}
\label{eq:12}
\left[\frac{\partial}{\partial t}+L^{(0)}(x_1)+L^{(0)}(x_2)\right]
g_2(x_1,x_2,t)
=\delta(\mathbf{r}_{12})\sigma^{d-1}\bar{T}_0(\mathbf{v}_1,\mathbf{v}_2)
f_1(x_1,t)f_1(x_2,t)\nonumber\\
+\left[K(x_1,t)+K(x_2,t)\right]g_2(x_1,x_2,t)
-\frac{\xi_0^2}{N}\frac{\partial}{\partial\mathbf{v}_1}\cdot
\frac{\partial}{\partial\mathbf{v}_2}f_1(x_1,t)f_1(x_2,t),
\end{eqnarray}
where we have introduced the linear operator $K(x_i,t)$ defined as
\begin{equation}
K(x_i,t)=\sigma^{d-1}\int dx_3\delta(\mathbf{r}_{i3})\bar{T}_0(\mathbf{v}_i,\mathbf{v}_3)
(1+{\cal P}_{i3})f_1(x_3,t)
+\frac{\xi_0^2}{2}\left(\frac{\partial}{\partial\mathbf{v}_i}\right)^2,
\end{equation}
and where the permutation operator ${\cal P}_{ab}$ interchanges the labels of 
particles $a$ and $b$ in the quantities on which it acts.
As will become clear below, the $1/N$ term in equation 
(\ref{eq:12}) is crucial for 
the calculation of the energy fluctuations.


\section{The stationary state}
\label{The stationary state}

It has been shown numerically that, after a transient time, the system reaches 
a homogeneous stationary state \cite{vnetp99} in which the energy input
from the thermostat is compensated by the energy lost in collisions. In this
section we will particularize the equations of the previous section to this
state, summarizing the results that are already known about the
one-particle distribution function and that are required for our theoretical
analysis.

The Boltzmann-Fokker-Planck equation 
(\ref{ec:boltzmann}) for the distribution function, $f_H(\mathbf{v}_1)$, in 
the stationary homogeneous state is
\begin{equation}\label{ec_b}
\frac{\xi_0^2}{2}\left(\frac{\partial}{\partial\mathbf{v}_1}\right)^2
f_H(\mathbf{v}_1)+J[f_H\vert f_H]=0.
\end{equation} 
It is convenient to introduce the scaled distribution function $\chi_H$
\begin{equation}
\label{scaling_f}
f_H(\mathbf{v})=\frac{n_H}{v_H^d}\chi_H\left({c}\right),
\end{equation}
where $n_H$ is the homogeneous density, 
$v_H=\left(\frac{2 T_H}{m}\right)^{1/2}$ is the thermal velocity
defined from the granular temperature 
\be
T_H= \frac{2}{d n_H} \int d\mathbf{v}\frac{1}{2}mv^2f_H(\mathbf{v}), 
\ee
and 
$\mathbf{c}=\frac{\mathbf{v}}{v_H}$ is the rescaled velocity.
The distribution function has been studied in reference \cite{vne98},
where an approximate expression for $\chi_H({c})$ was derived to second order in 
Sonine polynomials \cite{LL}
\begin{equation}\label{ec:sonine}
\chi_H({c})=\frac{e^{-c^2}}{\pi^{d/2}}
\left(1+a_2(\alpha)S^2_{d/2-1}(c^2)\right),
\end{equation}
with 
\begin{equation}
S^2_{d/2-1}(c^2)=\frac{1}{2}c^4-\frac{1}{2}(d+2)c^2+\frac{1}{8}d(d+2),
\end{equation}
and $a_2(\alpha)$ a coefficient related to the kurtosis of the function 
$\chi_H({c})$
\begin{equation}
\frac{d}{d+2}\frac{\langle c^4\rangle_H}{\langle c^2\rangle_H^2}=1+a_2(\alpha).
\end{equation}
An approximate expression for $a_2$ reads (see \cite{ms00,cdpt03} for a 
discussion on various possible approximations)
\begin{equation}\label{a2}
a_2=\frac{16(1-\alpha)(1-2\alpha^2)}
{73+56d-24\alpha d-105\alpha+30(1-\alpha)\alpha^2}.
\end{equation}
The expression for the temperature in the first Sonine approximation is 
\be\label{T_H}
T_H=m\left[\frac{d\xi_0^2\sqrt{\pi}}
{(1-\alpha^2)\Omega_dn_H\sigma^{d-1}}\right](1+{\cal O}(a_2)),
\ee
where $\Omega_d=2\pi^{d/2}/\Gamma(d/2)$ is the $d$-dimensional solid angle. 

We now turn to the equation for the correlation function, 
$g_{2,H}(x_1,x_2)$. It is convenient to introduce the  rescaled correlation 
function $\widetilde{g}_H$ through
\begin{equation}
g_{2,H}(x_1,x_2)=\frac{n_H}{\ell^dv_H^{2d}}
\widetilde{g}_H(\mathbf{l}_{12},\mathbf{c}_1,\mathbf{c}_2),
\end{equation}
where $\ell=(n_H\sigma^{d-1})^{-1}$ is proportional to the mean free path and
$\mathbf{l}=\mathbf{r}/\ell$. In these units,
the equation for the reduced function $\widetilde{g}_H$  reads
\begin{eqnarray}\label{ec:g2}
\left[\Lambda(\mathbf{c}_1)+\Lambda(\mathbf{c}_2)
-\mathbf{c}_{12}\cdot\frac{\partial}{\partial\mathbf{l}_{12}}\right]
\widetilde{g}_H(\mathbf{l}_{12},\mathbf{c}_1,\mathbf{c}_2)
\qquad\qquad\qquad\qquad\qquad\qquad\nonumber\\
=-\delta(\mathbf{l}_{12}) \bar{T}_0(\mathbf{c}_1,\mathbf{c}_2)
\chi_H({c}_1)\chi_H({c}_2)+\widetilde{\xi}_0^2\frac{n_H\ell^d}{N}
\frac{\partial}{\partial\mathbf{c}_1}\cdot\frac{\partial}{\partial\mathbf{c}_2}
\chi_H({c}_1)\chi_H({c}_2),
\end{eqnarray}
where we have introduced the linearized Boltzmann-Fokker-Planck
operator 
$\Lambda(\mathbf{c})$
\begin{equation}\label{ec:lambda}
\Lambda(\mathbf{c}_i)h(\mathbf{c}_i)=
\int d\mathbf{c}_3\bar{T}_0(\mathbf{c}_i,\mathbf{c}_3)(1+P_{i3})\chi_H({c}_3)
h(\mathbf{c}_i)
+\frac{\widetilde{\xi}_0^2}{2}
\left(\frac{\partial}{\partial\mathbf{c}_i}\right)^2h(\mathbf{c}_i),
\end{equation}
with rescaled noise amplitude
\begin{equation}
\widetilde{\xi}_0^2=\frac{\xi_0^2 l}{v_H^3}.
\end{equation}

As can be seen in equation (\ref{ec:g2}), the correlation
function, $\widetilde{g}_H$, is determined by the properties of the linearized
Boltzmann-Fokker-Planck operator, $\Lambda$, and by the one particle distribution function
$\chi_H$. It is consequently important to study the spectral
properties of $\Lambda$, in particular the upper 
(hydrodynamic) part of the spectrum,
in order to understand the fluctuations of global 
quantities. In the case of a granular gas in the homogeneous cooling
state \cite{bdr03}, and 
for a system under ballistic annihilation dynamics  \cite{gmsbt08}, it has been 
shown that it is possible to find the hydrodynamic eigenvalues and 
eigenfunctions of the linearized Boltzmann-Fokker-Planck operator. 
Once these quantities are known, it becomes possible 
to evaluate the
fluctuations of the relevant global quantities 
in the so-called ``hydrodynamic approximation''
\cite{bgmr04,mgsbt08}. In the remainder, 
we will see that we can evaluate the fluctuations of the total energy in
an equivalent approximation, but without
the knowledge of the eigenfunction associated to the energy. The only
information needed is the form of the linearized hydrodynamic equations and,
in particular, the eigenvalues.


\section{Hydrodynamic equations}
\label{Hydrodynamic equations}

\subsection{Evolution of homogeneous perturbations}
\label{ssec:hom}
In this section we focus on the linearized
hydrodynamic equations around the homogeneous stationary state. The 
objective 
is to consider the linearized equations around a homogeneous perturbation in
order to extract information about the linear behavior of a small perturbation
of the total energy. 

The complete non-linear hydrodynamic equations for the
granular system heated by the stochastic thermostat are \cite{vnetp99,gm02}
\begin{eqnarray}
\frac{\partial}{\partial t}n&=&-\nabla\cdot(n\mathbf{u}),\\
\frac{\partial}{\partial t}\mathbf{u}
&=&-\mathbf{u}\cdot\nabla\mathbf{u}-\frac{1}{mn}\nabla_jP_{ij},\\
\frac{\partial}{\partial t}T&=&-\mathbf{u}\cdot\nabla T
-\frac{2}{dn}(\nabla\cdot\mathbf{q}+P_{ij}\nabla_ju_i)-\zeta T+m\xi_0^2,
\end{eqnarray}
where $P_{ij}$ is the pressure tensor, $\mathbf{q}$ is the heat flux and 
$\zeta$ is the cooling rate, which is also a functional of the distribution 
function
\begin{equation}
\zeta=\frac{(1-\alpha^2)m\pi^{\frac{d-1}{2}}\sigma^{d-1}}
{4d\Gamma\left(\frac{d+3}{2}\right)nk_BT}
\int d\mathbf{v}_1\int d\mathbf{v}_2
\vert\mathbf{v}_1-\mathbf{v}_2\vert^3
f(\mathbf{r},\mathbf{v}_1,t)f(\mathbf{r},\mathbf{v}_2,t). 
\end{equation}
Considering a homogeneous state, the previous equations reduce to
\be\label{homogeneous_ec}
\frac{\partial}{\partial t}n=0,\qquad
\frac{\partial}{\partial t}\mathbf{u}=\mathbf{0},\qquad
\frac{\partial}{\partial t}T=-\zeta T+m\xi_0^2.
\ee
In the long time limit, the system is expected to approach a steady state with 
a constant temperature given by the equation
\be\label{homogeneous_ec2}
\zeta_H(f_H)T_H=m\xi_0^2.
\end{equation}
Substituting the explicit form of the one particle distribution function 
(\ref{scaling_f}) in the equation above, we obtain the temperature given in
equation (\ref{T_H}).

Let us consider now a homogeneous state close to this homogeneous stationary 
state. We can write the hydrodynamic fields as $n(t)=n_H+\delta n$, 
$\mathbf{u}(t)=\delta \mathbf{u}$ and $T(t)=T_H+\delta T$. We also define 
the dimensionless hydrodynamic fields
\begin{equation}
\delta\rho(\tau)=\frac{\delta n}{n_H},\qquad
\delta\mathbf{w}(\tau)=\frac{\delta \mathbf{u}}{v_H},\qquad
\delta \theta(\tau)=\frac{\delta T}{T_H}, 
\end{equation}
where we have introduced the dimensionless time scale $\tau$, proportional to
the number of collisions per particle, defined as
\be
\tau=\int_0^t dt'\frac{v_H}{\ell}=\frac{v_H}{\ell}t. 
\ee
Assuming that the deviations are small, and taking into account equations 
(\ref{homogeneous_ec})-(\ref{homogeneous_ec2}), we can write the linearized
evolution equations for the dimensionless hydrodynamic fields in this new time 
scale 
\begin{equation}\label{ec:desviaciones}
\frac{\partial}{\partial \tau}\delta \rho=0,\qquad
\frac{\partial}{\partial \tau}\delta \mathbf{w}=0,\qquad
\frac{\partial}{\partial \tau}\delta \theta=-\zeta_0\delta\rho
-\frac{3}{2}\zeta_0\delta\theta,
\end{equation}
where $\zeta_0=\frac{l\zeta_H}{v_H}$ is a dimensionless coefficient that
  is a functional of the one-particle distribution function in the stationary
  state. Its expression in 
  the first Sonine approximation is \cite{vne98}
\begin{equation}\label{zeta0}
\zeta_0=\frac{(16+3a_2) \pi^{\frac{d-1}{2}} (1-\alpha^2)}{
 8 \sqrt{2} d \Gamma\left(\frac{d}{2}\right)}.
\end{equation}
To obtain the equation for $\delta \theta$ we have assumed that the perturbed
distribution function scales as
\be
f(\mathbf{v},t)=\frac{n}{\bar{v}(t)^d}
\chi_H\left(\frac{\mathbf{v}}{\bar{v}(t)}\right), 
\ee
where $\bar{v}(t)=\left[\frac{2T(t)}{m}\right]^{1/2}$, and $\chi_H$ is the same
scaled distribution function as for the reference stationary
state. This assumption has already been used and tested numerically in
\cite{vnetp99}. Then, the 
cooling rate $\zeta$ for the state under scrutiny is proportional to 
$T^{1/2}(t)$ and we obtain the equation for the linearized energy written
above. Equations (\ref{ec:desviaciones}) indicate that a perturbation in the 
total number of particles or total momentum does not decay, as a consequence of the fact that
these variables are conserved, but a perturbation in the total energy will
decay (exponentially in $\tau$) to the stationary value, as
expected. Moreover, as the
equation for the temperature can be rewritten in the following form
\be
\frac{\partial}{\partial \tau}\left(\frac{2}{3}\delta\rho+\delta\theta\right)
=-\frac{3}{2}\zeta_0\left(\frac{2}{3}\delta\rho+\delta\theta\right), 
\ee
we can identify the hydrodynamic eigenvalues  $\lambda=0$ and 
$\gamma=-\frac{3}{2}\zeta_0$, $\lambda$ being $(d+1)$-fold degenerate. 
For the sake of clarity, 
it proves convenient to relabel these eigenvalues as 
\begin{equation}\label{eigenvalues}
\lambda_1=0,\qquad\lambda_2=0,\qquad\lambda_3=-\frac{3}{2}\zeta_0,
\end{equation} 
where $\lambda_2$ is $d$-fold degenerate. The associated hydrodynamic
modes, $\{y_\beta\}_{\beta=1}^{d+2}$ are
\be\label{hidro_modes}
y_1=\delta\rho, \qquad \mathbf{y}_2=\delta\mathbf{w}, \qquad 
y_3=\frac{2}{3}\delta\rho+\delta\theta. 
\ee
The $+$ sign in the last equation stems from the fact that an 
increased density leads to enhanced dissipation, and hence,
a smaller temperature.

\subsection{Enforcing consistency with the linearized Boltzmann-Fokker-Planck equation description}
We now turn our attention to the problem of finding the linearized hydrodynamic
equations for a homogeneous perturbation, directly from the 
Boltzmann-Fokker-Planck
equation. Enforcing consistency with the macroscopic considerations
of section \ref{ssec:hom} above, we will infer useful properties
on the 
hydrodynamic part of the spectrum of $\Lambda(\mathbf{c})$. We first 
introduce
the scaled deviation of the distribution function 
\be
\delta\chi(\mathbf{c},\tau)
=\frac{v_H^d}{n_H}[f(\mathbf{v},t)-f_H(\mathbf{v})]. 
\ee
The evolution of the scaled distribution is governed by 
\be\label{ec_B_L}
\frac{\partial}{\partial\tau}\delta\chi(\mathbf{c},\tau)
=\Lambda({\mathbf{c}})\delta\chi(\mathbf{c},\tau), 
\ee
where the operator $\Lambda(\mathbf{c})$ is the linearized Boltzmann-Fokker-Planck operator
defined in (\ref{ec:lambda}). Let us also introduce the scalar product 
\begin{equation}
\langle f(\mathbf{c})\vert g(\mathbf{c})\rangle
\equiv\int d\mathbf{c}\chi_H^{-1}({c})f^*({\mathbf{c}})g({\mathbf{c}}),
\end{equation}
where $f^*$ is the complex conjugate of $f$. Interestingly, the hydrodynamic modes
introduced in (\ref{hidro_modes}) can then be written as
\be
y_\beta=\langle\bar{\xi}_\beta\vert\delta\chi\rangle, \qquad \beta=1,2,3, 
\ee
where 
\be\label{bar_xi}
\bar{\xi}_1(\mathbf{c})=\chi_H({c}), \qquad 
\boldsymbol{\bar{\xi}}_2(\mathbf{c})=\mathbf{c}\chi_H({c}),\qquad
\bar{\xi}_3(\mathbf{c})=\left(\frac{c^2}{d}-\frac{1}{6}\right)\chi_H({c}).
\ee
Taking the scalar product of the linearized Boltzmann-Fokker-Planck equation 
(\ref{ec_B_L}) with the  functions $\bar{\xi}_\beta$, we obtain the linear
equations (\ref{ec:desviaciones}) (in the hydrodynamic time scale,
that is, if we wait long enough so that fast modes have vanished) only if the
spectrum of $\Lambda$ admits the three eigenvalues written in 
(\ref{eigenvalues}), and the associated ``hydrodynamic'' eigenfunctions,
$\{\xi_\beta\}_{\beta=1}^{d+2}$, obey the orthogonality condition
\begin{equation}\label{biorth_cond}
\langle\bar{\xi}_{\beta_1}\vert\xi_{\beta_2}\rangle=\delta_{\beta_1\beta_2},
\qquad\beta_1,\beta_2=1,2,3. 
\end{equation} 
In Appendix \ref{apendicea},  it is shown that the null eigenvalue is
$(d+1)$-fold degenerate, and the 
corresponding eigenfunctions, $\xi_1$ and $\boldsymbol{\xi}_2$ , 
are worked out. Moreover, 
as a consequence of particle and total momentum conservation in a collision, 
$\bar{\xi_1}$ and $\boldsymbol{\bar{\xi_2}}$ are the
corresponding left eigenfunctions. We were not able to demonstrate that
$\lambda_3$ is an eigenvalue of $\Lambda$, but we have shown explicitly
that 
\be
\langle\bar{\xi}_{3}\vert\xi_{\beta}\rangle=0, \qquad \text{for} 
\qquad \beta=1,2. 
\ee
In the following, we will assume that $\Lambda$ actually admits  this third
eigenvalue, with an unknown eigenfunction $\xi_3$. With the help of this
assumption, we will see in the next section that it is possible to define a
projector in the hydrodynamic subspace, which opens the way for evaluating
the variance of the total
energy fluctuations.


\section{Energy fluctuations}
\label{Energy fluctuations}

In this section, we study the fluctuations of the global energy for a system in 
the stationary state. As we are interested in global quantities, it is 
convenient to define a global correlation function $\phi_H$
\begin{equation}
\phi_H(\mathbf{c}_1,\mathbf{c}_2)\equiv\int d\mathbf{r}_{12}
\widetilde{g}_H(\mathbf{r}_{12},\mathbf{c}_1,\mathbf{c}_2).
\end{equation}
The energy fluctuations can be written as a functional of this correlation
function and the one-particle distribution function $\chi_H$, as
\cite{bgmr04,gmsbt08} 
\be\label{deltaE}
\langle(\delta E)^2\rangle_H=\frac{m^2}{4}Nv_H^4
\left[\int d\mathbf{c}c^4\chi_H(c)
+\int d\mathbf{c}_1\int d\mathbf{c}_2 c_1^2c_2^2
\phi_H(\mathbf{c}_1,\mathbf{c}_2)\right]. 
\ee
In order to evaluate
the integral over the correlation function $\phi_H$, 
we start from (\ref{ec:g2}), integrating over the position variable. 
Assuming periodic boundary conditions, the spacial gradient terms disappear
and we have the following equation for $\phi_H$
\begin{equation}\label{ec:g2esc}
\left[\Lambda(\mathbf{c}_1)+\Lambda(\mathbf{c}_2)\right]
\phi_H(\mathbf{c}_1,\mathbf{c}_2)
=\Gamma(\mathbf{c}_1,\mathbf{c}_2),
\end{equation}
where
\begin{equation}\label{gamma}
\Gamma(\mathbf{c}_1,\mathbf{c}_2)=
-\bar{T}_0(\mathbf{c}_1,\mathbf{c}_2)\chi_H({c}_1)\chi_H({c}_2)
+\widetilde{\xi}_0^2
\frac{\partial}{\partial\mathbf{c}_1}\cdot\frac{\partial}{\partial\mathbf{c}_2}
\chi_H({c}_1)\chi_H({c}_2). 
\end{equation}
The solubility condition for equation (\ref{ec:g2esc}) is that 
$\Gamma$ does not have components in the subspace associated to the null
eigenvalue. In our case, this subspace is generated by 
$\{\bar{\xi}_1,\boldsymbol{\bar{\xi}_2}\}$. Due to the conservation of the
number of particles and total momentum in a collision, and to the symmetry of
the second term of $\Gamma$, we have 
\begin{eqnarray}
\langle\bar{\xi}_1(\mathbf{c}_1)\boldsymbol{\bar{\xi}_2(\mathbf{c}_2)}
\vert\Gamma(\mathbf{c}_1,\mathbf{c}_2)\rangle=
\langle\boldsymbol{\bar{\xi}_2(\mathbf{c}_1)}\bar{\xi}_1(\mathbf{c}_2)
\vert\Gamma(\mathbf{c}_1,\mathbf{c}_2)\rangle=\mathbf{0},\\
\langle\bar{\xi}_1(\mathbf{c}_1)\bar{\xi}_1(\mathbf{c}_2)
\vert\Gamma(\mathbf{c}_1,\mathbf{c}_2)\rangle=
\langle\bar{\xi}_{2,i}(\mathbf{c}_1)\bar{\xi}_{2,j}(\mathbf{c}_2)
\vert\Gamma(\mathbf{c}_1,\mathbf{c}_2)\rangle=0,
\end{eqnarray}
for $i\ne j$. The case $i=j$ is analyzed in the Appendix \ref{apendiceb}, 
where it is shown that $\langle\bar{\xi}_{2,i}(\mathbf{c}_1)
\bar{\xi}_{2,i}(\mathbf{c}_2)\vert\Gamma(\mathbf{c}_1,\mathbf{c}_2)\rangle=0$.
In order to prove this property, the presence of the second term of the
  right hand side of Eq. (\ref{gamma}) is essential. Hence, the solubility
condition holds and the problem of finding $\phi_H$ with equation 
(\ref{ec:g2esc}) is well defined.

\begin{figure}
\begin{center}
\includegraphics[angle=0,width=0.7\textwidth]{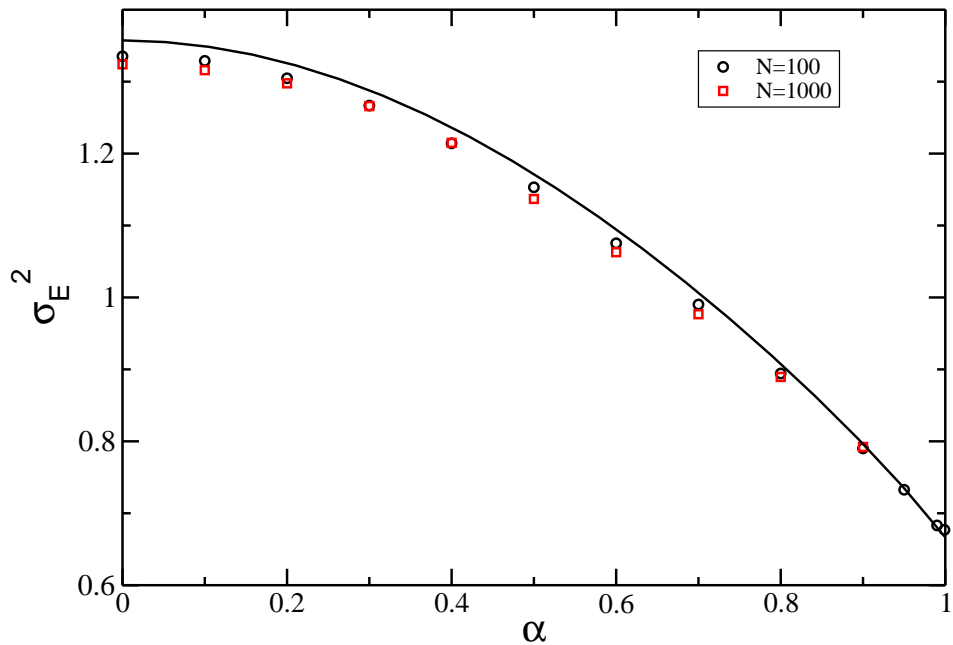}
\end{center}
\caption{Scaled second moment of the energy fluctuations $\sigma_E^2$ as a 
function of the restitution coefficient $\alpha$. The solid line is the 
theoretical prediction and symbols are the two dimensional 
Monte Carlo simulation results
of Ref. \cite{vpbvwt06}.}
\label{figura1}
\end{figure}

Let us also define a projector $P_{12}$ in the hydrodynamic subspace as 
\begin{equation}
P_{12} h(\mathbf{c}_1,\mathbf{c}_2)=\sum_{\beta_1=1}^3\sum_{\beta_2=1}^3 
\langle\bar{\xi}_{\beta_1}(\mathbf{c}_1)\bar{\xi}_{\beta_2}(\mathbf{c}_2)
\vert h(\mathbf{c}_1,\mathbf{c}_2)\rangle 
\xi_{\beta_1}(\mathbf{c}_1)\xi_{\beta_2}(\mathbf{c}_2),
\end{equation}
where $\{\xi_\beta\}_{\beta=1}^3$ are the right hydrodynamic eigenfunctions of 
the linearized Boltzmann-Fokker-Planck operator, $\{\bar{\xi}_\beta\}_{\beta=1}^3$ the
orthogonal set introduced in the previous section, equation (\ref{bar_xi}), 
and we have generalized the scalar product by
\begin{equation}
\langle f(\mathbf{c}_1,\mathbf{c}_2)\vert g(\mathbf{c}_1,\mathbf{c}_2)\rangle
=\int d\mathbf{c}_1\int d\mathbf{c}_2
\chi_H^{-1}({c}_1)\chi_H^{-1}({c}_2)
f^*(\mathbf{c}_1,\mathbf{c}_2) g(\mathbf{c}_1,\mathbf{c}_2).
\end{equation}
Note that $P_{12}$ is a projector even if the function $\bar{\xi}_3$ is not
the true left eigenfunction of $\Lambda(\mathbf{c})$  (remember that 
$\bar{\xi}_1$ and $\bar{\boldsymbol{\xi}}_2$ are the actual left
eigenfunctions associated to the null eigenvalue). The fact that the set of
functions $\{\bar{\xi}_\beta\}_{\beta=1}^3$ and 
$\{\xi_\beta\}_{\beta=1}^3$ fulfil the orthogonality condition
(\ref{biorth_cond}) is enough to guarantee that 
$P^2_{12}=P_{12}$. Using this projector, we define the ``hydrodynamic part'' 
of $\phi_H$ to be the function
\begin{equation}\label{ec:descomp}
\phi_H^{(h)}(\mathbf{c}_1,\mathbf{c}_2)\equiv 
P_{12}\phi_H(\mathbf{c}_1,\mathbf{c}_2)=
\sum_{\beta_1=1}^3\sum_{\beta_2=1}^3 a_{\beta_1\beta_2}
\xi_{\beta_1}(\mathbf{c}_1)\xi_{\beta_2}(\mathbf{c}_2). 
\end{equation}
The coefficients $a_{\beta_1\beta_2}$ are the quantities we need to evaluate. As
they are essentially the first moments of the correlation function $\phi_H$,
they are directly related to the integral we have to calculate
in (\ref{deltaE}). It is tempting to treat $\xi_3$ as if it was an actual
left eigenfunction of the linearized Boltzmann-Fokker-Planck operator, and we
will in the following use the approximation
\begin{equation}\label{ec:aprox}
P_{12}\Lambda(\mathbf{c}_i)=P_{12}\Lambda(\mathbf{c}_i) P_{12}, 
\end{equation}
which allows to
find a closed equation for $\phi_H^{(h)}$.
This approximation has already been invoked in
  other systems such as the freely evolving granular gas \cite{bgmr04}, or the 
  probabilistic ballistic annihilation model \cite{mgsbt08}. With the
  information available on the linearized Boltzmann-Fokker-Planck operator,
  it seems the best that can be done technically. Let us also remark that the approximation is exact
  in the elastic limit. Then, applying the projector $P_{12}$ to the equation 
(\ref{ec:g2esc}) and taking into account the approximation (\ref{ec:aprox}),
we obtain the following expressions for the coefficients $a_{\beta_1\beta_2}$ 
\begin{equation}\label{ec:coef}
a_{\beta_1\beta_2}=-\frac{\langle\bar{\xi}_{\beta_1}(\mathbf{c}_1)
\bar{\xi}_{\beta_2}(\mathbf{c}_2)\vert \bar{T}_0(\mathbf{c}_1,\mathbf{c}_2)
\chi_H({c}_1)\chi_H({c}_2)\rangle}{\lambda_{\beta_1}+\lambda_{\beta_2}},
\end{equation}
where it has been assumed that $\lambda_{\beta_1}+\lambda_{\beta_2}\ne 0$. The
coefficients associated to the vanishing eigenvalue cannot be calculated by
equation (\ref{ec:g2esc}), but are fixed by the boundary conditions. In  
Appendix \ref{apendicec}, the coefficients $a_{\beta_1\beta_2}$ are 
evaluated. The expression for $\phi_H^{(h)}$ is finally given by
\begin{equation}\label{phiHh}
\phi_H^{(h)}=a_{11}\xi_1(\mathbf{c}_1)\xi_1(\mathbf{c}_2)
+a_{13}\left[\xi_1(\mathbf{c}_1)\xi_3(\mathbf{c}_2)
+\xi_3(\mathbf{c}_1)\xi_1(\mathbf{c}_2)\right]
+a_{33}\xi_3(\mathbf{c}_1)\xi_3(\mathbf{c}_2),
\end{equation}
where $a_{11}=-1$, $a_{13}=-\frac{1}{3}$ and $a_{33}$ can be obtained as a
functional of the one-particle distribution function
\begin{equation}
a_{33}=\frac{\langle\bar{\xi}_3(\mathbf{c}_1)\bar{\xi}_3(\mathbf{c}_2)
\vert\bar{T}_0(\mathbf{c}_1,\mathbf{c}_2)\chi_H(c_1)\chi_H(c_2)\rangle}
{3\zeta_0}.
\end{equation}
An approximate expression is derived in appendix \ref{apendicec},
and reads 
\begin{equation}
a_{33}=\frac{-15+7d+14d^2-3(-9+d(9+2d))\alpha+30(1+d)\alpha^2-6(9+d)\alpha^3)}
{9d(-19+2d(-7+3\alpha)+3\alpha(9+2(-1+\alpha)\alpha))}.
\end{equation}
Taking into account (\ref{deltaE}) and (\ref{phiHh}),
the variance of the energy fluctuations can finally be written as
\be\label{ec:sigma}
\sigma_E^2=N\frac{\langle(\delta E)^2\rangle_H}{\langle E\rangle_H^2}
=(a_2+1)\frac{d(d+2)}{4}+d^2a_{33}-d^2\frac{5}{36}. 
\ee
In reference \cite{vpbvwt06}, the value of $\sigma_E^2$ has been measured by 
means of the Direct Monte Carlo simulation method (DSMC).
In Fig.~\ref{figura1}, we compare our theoretical prediction (solid line given 
by (\ref{ec:sigma})) with the DMSC simulations results (symbols). The 
agreement is satisfactory for the whole range of inelasticities,
at variance with the theoretical attempt put forward in \cite{vpbvwt06},
which neglected velocity correlations.
In particular, we note the non-trivial result in the elastic limit
$\lim_{\alpha\to 1^{-}}\sigma_E^2(\alpha)=d/3$
(i.e. 2/3 on Fig. \ref{figura1}, which is well obeyed), while
in the free cooling regime, this quantity vanishes \cite{bgmr04}.
We emphasize that the elastic limit is singular: 
the behaviour for elastic systems with $\alpha=1$ is not approached
by taking the quasi-elastic limit $\alpha \to 1^-$ (we note that the divergence of the different moments of the velocity
distribution as $\alpha\to 1$ is nevertheless indicative of the absence
of a stationary state when $\alpha=1$). Such a singularity
has already been reported in 1 dimension \cite{02}, but, to 
our knowledge, not for two dimensional granular systems. It is also
interesting to note that the singular nature of 
the quasi-elastic limit
appears at 2 body level through the energy fluctuations, 
while, as far as rescaled distribution functions
are considered, the 1 body level of description is regular, with a well
behaved velocity distribution approaching a Gaussian form \cite{ptvne01}.

\section{Conclusions}

The problem of the fluctuations of the total energy of a granular (inelastic) gas 
maintained in a non-equilibrium stationary state by a random acceleration has 
been addressed. A numerical study of this quantity had been performed by means of 
Monte Carlo simulations and an argument assuming  uncorrelated non-Gaussian individual 
distribution function had been proposed in \cite{vpbvwt06}, without success. 
The main goal of 
this work was therefore to take due account of velocity 
correlations in order to study
these fluctuations. 

To this end, the standard description at the single particle level
is not sufficient, and
the two-particle correlation function is needed. We have 
derived the evolution equation for such an object, 
and particularized the analysis to the homogeneous 
stationary state that is reached by the system in the long time limit. 
Our work shows that this equation is not a straightforward 
generalization of its counterpart arising in the context
of the undriven granular gas
(i.e. by only changing the linearized Boltzmann-Fokker-Planck operator 
into its driven form). 
A non trivial non diagonal term appears in the Fokker-Planck equation for the two
particle distribution function (contribution $\partial_{\mathbf{v}_1}\cdot
\partial_{\mathbf{v}_2}$ in Eq. (\ref{eq:bfp})),
 as a consequence of the coupling between velocities 
 due to momentum conservation.

We have seen that, for our purposes, the exact knowledge of the hydrodynamic 
eigenfunctions is not needed. The important point is that we can construct a
set of functions $\{\bar{\xi}_\beta\}_{\beta=1}^3$, which are linear 
combinations of $1$, $\mathbf{c}$ and $c^2$, that are orthogonal to the
right eigenfunctions $\{\xi_\beta\}_{\beta=1}^3$ of the
linearized Boltzmann-Fokker-Planck operator $\Lambda$. This orthogonality property 
holds for the ``real'' left eigenfunctions, that in our case correspond to 
the null eigenvalue (i.e. density and velocity fields
associated to conserved quantities). The function $\bar{\xi}_3$ 
is not a left eigenfunction of $\Lambda$ but it
can be proved to be orthogonal to $\xi_1$ and $\boldsymbol{\xi}_2$. 
In a subsequent step, 
 the linear hydrodynamic equations around the reference state are 
derived and from that knowledge, 
the hydrodynamic eigenvalues are identified and the  variance of
energy fluctuations
subsequently derived.





Finally, our prediction has  been successfully tested against
the numerical results obtained by the Direct Monte Carlo simulation method 
for all the range of values of the coefficient
of normal restitution $\alpha$. This provides strong support for the theory
developed here and assesses in retrospect the validity of our assumptions.

\begin{acknowledgments}
It is a pleasure to dedicate this work to Jean-Jacques Weis.
We thank Paolo Visco for useful discussions, and for providing us with the
Monte Carlo data of Fig. \ref{figura1}. We would like to thank the Agence
Nationale de la Recherche for financial support (grant ANR-05-JCJC-44482). 
M.~I.~G.~S.~ and P. M. acknowledge financial support from Becas de la
Fundaci\'on La Caixa y el Gobierno Franc\'es. E.T. acknowledges the support of
Institut Universitaire de France.

\end{acknowledgments}


\appendix

\section{Eigenvalue problem for $\Lambda$}\label{apendicea}

We consider here the eigenvalue problem for the 
homogeneous linear Boltzmann-Fokker-Planck operator $\Lambda$, defined in (\ref{ec:lambda})
\begin{equation}
\Lambda(\mathbf{c})\xi_\beta(\mathbf{c})=\lambda_\beta\xi_\beta(\mathbf{c}).
\end{equation}
We are interested in the eigenfunctions and eigenvalues associated 
with linear hydrodynamics and, to perform the analysis, 
similar techniques as in \cite{bdr03,bgmr04,gmsbt08} will be required.

Consider first the function
\begin{equation}
\psi_1(\mathbf{c})=\chi_H({c}).
\end{equation}
When the linearized operator $\Lambda$ acts on $\chi_H$, we have
\begin{equation}
\Lambda(\mathbf{c}_1)\chi_H(\mathbf{c}_1)=\int d\mathbf{c}_2 
\bar{T}_0(\mathbf{c}_2,\mathbf{c}_3)(1+{\cal P}_{12})
\chi_H({c}_2)\chi_H({c}_1)
+\frac{\widetilde{\xi}_0^2}{2}\left(\frac{\partial}
{\partial\mathbf{c}_1}\right)^2\chi_H({c}_1).
\end{equation}
Taking into account the equation for $\chi_H$, Ec. (\ref{ec_b}), we obtain the
following relation
\begin{equation}\label{ec:psi1}
\Lambda(\mathbf{c}_1)\psi_1(c_1)=-\frac{\widetilde{\xi}_0^2}{2}
\left(\frac{\partial}{\partial\mathbf{c}_1}\right)^2\chi_H(c_1).
\end{equation}

Now, let us considerer the function
\begin{equation}
\boldsymbol{\psi}_2(\mathbf{c})=-\frac{\partial}
{\partial\mathbf{c}}\chi_H(c).
\end{equation}
Taking derivate in the equation obeyed by $\chi_H(\mathbf{c}-\mathbf{w})$ with 
respect to $\mathbf{w}$, and subsequently evaluating the result for 
$\mathbf{w}=0$, we obtain
\begin{equation}\label{ec:psi2}
\Lambda(\mathbf{c}_1)\boldsymbol{\psi}_2(\mathbf{c}_1)=\mathbf{0}.
\end{equation}

Finally, we will consider the function
\begin{equation}
\psi_3(\mathbf{c})
=\mathbf{c}\cdot\frac{\partial}{\partial\mathbf{c}}\chi_H(c).
\end{equation}
From the equation obeyed by $\psi_3(\lambda\mathbf{c}_1)$, we can
take derivate with respect to $\lambda$, and evaluate the result for 
$\lambda=1$. We arrive at an equation for $\psi_3(\mathbf{c}_1)$,
\begin{equation}\label{ec:psi3}
\Lambda(\mathbf{c}_1)\psi_3(\mathbf{c}_1)=(d+3)\frac{\widetilde{\xi}_0^2}{2}
\left(\frac{\partial}{\partial\mathbf{c}_1}\right)^2\chi_H(c_1).
\end{equation}

From equations (\ref{ec:psi1}), (\ref{ec:psi2}) and (\ref{ec:psi3}), we can 
identify two eigenfunctions of $\Lambda$. Making use of (\ref{ec:psi1}) and 
(\ref{ec:psi3}), it appears that
\begin{equation}\label{ec:psi3b}
\Lambda(\mathbf{c})\left(\frac{1}{3}\frac{\partial}{\partial\mathbf{c}}
\cdot\left[\mathbf{c}\chi_H(c)\right]+\chi_H(c)\right)=0.
\end{equation}

 Hence, from Eqs. (\ref{ec:psi2}) and (\ref{ec:psi3b}) we can conclude that 
the null eigenvalue is $(d+1)$-fold degenerate with the eigenfunctions
\begin{equation}
\xi_1(\mathbf{c})=\frac{1}{3}\frac{\partial}
{\partial\mathbf{c}}\cdot\left[\mathbf{c}\chi_H(c)\right]+\chi_H(c), \qquad
\boldsymbol{\xi}_2=-\frac{\partial}{\partial\mathbf{c}}\chi_H(c).
\end{equation}

\section{Evaluation of the coefficient
$a_{2,i2,i}$}\label{apendiceb}

 In this appendix, we  show $\langle\bar{\xi}_{2,i}(\mathbf{c}_1)
\bar{\xi}_{2,i}(\mathbf{c}_2)\vert\Gamma(\mathbf{c}_1,\mathbf{c}_2)\rangle=0$.
The integral corresponding to the second term of $\Gamma$ is simply
\be
\int d\mathbf{c}_1\int d\mathbf{c}_2c_{1,i}c_{2,i}\widetilde{\xi}_0^2
\frac{\partial}{\partial\mathbf{c}_1}\cdot\frac{\partial}{\partial\mathbf{c}_2}
\chi_H({c}_1)\chi_H({c}_2)=\widetilde{\xi}_0^2. 
\ee
The other term can be written as
\begin{eqnarray}
&&\int d\mathbf{c}_1\int d\mathbf{c}_2c_{1,i}c_{2,i}
\bar{T}_0(\mathbf{c}_1,\mathbf{c}_2)\chi_H({c}_1)\chi_H({c}_2)\nonumber\\
&&\qquad=\int d\mathbf{c}_1\int d\mathbf{c}_2
\chi_H({c}_1)\chi_H({c}_2)\int d\boldsymbol{\hat{\sigma}}
\Theta(\boldsymbol{\hat{\sigma}}\cdot\mathbf{c}_{12})
(\boldsymbol{\hat{\sigma}}\cdot\mathbf{c}_{12})
\left[b_{\boldsymbol{\hat{\sigma}}}-1\right]c_{1,i}c_{2,i}\nonumber\\
&&\qquad=\int d\mathbf{c}_1\int d\mathbf{c}_2
\chi_H({c}_1)\chi_H({c}_2)\int d\boldsymbol{\hat{\sigma}}
\Theta(\boldsymbol{\hat{\sigma}}\cdot\mathbf{c}_{12})
\left[(\boldsymbol{\hat{\sigma}}\cdot\mathbf{c}_{12})
\frac{1+\alpha}{2}(\boldsymbol{\hat{\sigma}}\cdot\mathbf{c}_{12})
c_{12,i}\hat{\sigma}_i\right.\nonumber\\
&&\qquad\qquad\qquad\qquad\qquad\qquad\qquad\left.-\frac{(1+\alpha)^2}{4}
(\boldsymbol{\hat{\sigma}}\cdot\mathbf{c}_{12})^2\hat{\sigma}_i^2
\right]\nonumber\\ 
&&\qquad=\frac{\pi^{(d-1)/2}}{\Gamma\left(\frac{d+3}{2}\right)}
\frac{1-\alpha^2}{4d}\int d\mathbf{c}_1\int d\mathbf{c}_2
\chi_H({c}_1)\chi_H({c}_2)c_{12}^3=\widetilde{\xi}_0^2,
\end{eqnarray}
which is the desired result.

\section{Evaluation of the coefficients  
$a_{\beta\beta^\prime}$}\label{apendicec}

In this appendix we evaluate the coefficients $a_{\beta\beta^\prime}$. As the 
number of particles and the total momentum are conserved quantities in our 
system, we have
\begin{eqnarray}
&\langle(\delta N)^2\rangle=0,\qquad\langle\delta P_i\delta P_j\rangle=0,\\
&\langle\delta N\delta P_i\rangle=0,\qquad\langle\delta N\delta E\rangle=0,\\
&\langle\delta E\delta P_i\rangle=0.
\end{eqnarray} 
Enforcing the above constraints, we obtain  
\begin{eqnarray}
\int d\mathbf{c}\chi_H(c)
+\int d\mathbf{c}_1\int d\mathbf{c}_2 
\phi_H(\mathbf{c}_1,\mathbf{c}_2)&=&1+a_{11}=0,\\
\int d\mathbf{c} c_i\chi_H(c)
+\int d\mathbf{c}_1\int d\mathbf{c}_2 c_{1i} 
\phi_H(\mathbf{c}_1,\mathbf{c}_2)&=&a_{12}=0,\\
\int d\mathbf{c} c^2\chi_H(c)
+\int d\mathbf{c}_1\int d\mathbf{c}_2 c_1^2\phi_H(\mathbf{c}_1,\mathbf{c}_2)
&=&\frac{d}{2}+da_{13}+\frac{d}{6}a_{11}=0,\\
\int d\mathbf{c} c_i c_j\chi_H(c)
+\int d\mathbf{c}_1\int d\mathbf{c}_2 c_{1i}c_{2j}
\phi_H(\mathbf{c}_1,\mathbf{c}_2)
&=&\frac{1}{2}\delta_{ij}+a_{2i2j}=0,\\
\int d\mathbf{c}c_ic^2\chi_H(c)
+\int d\mathbf{c}_1\int d\mathbf{c}_2 c_{1i}c_2^2
\phi_H(\mathbf{c}_1,\mathbf{c}_2)&=&a_{23}=0.
\end{eqnarray}
As a consequence, the values of some coefficients follow
\begin{equation}
a_{11}=-1,\quad a_{12}=0,\quad a_{13}=-\frac{1}{3},\quad 
a_{2i2j}=-\frac{1}{2}\delta_{ij},\quad a_{23}=0.
\end{equation}
Of course, the coefficients associated to 
$\lambda_{\beta_1}+\lambda_{\beta_1}\ne 0$ could also have been calculated
directly by equation (\ref{ec:coef}), obtaining the same results. 
The coefficient $a_{33}$ is evaluated using (\ref{ec:coef}) and it can be 
written in terms of the one-particle distribution function as
\begin{eqnarray}
a_{33}&=&\frac{\langle\bar{\xi}_3(\mathbf{c}_1)\bar{\xi}_3(\mathbf{c}_2)
\vert\bar{T}_0(\mathbf{c}_1,\mathbf{c}_2)\chi_H(c_1)\chi_H(c_2)\rangle}
{3\zeta_0}\nonumber\\
&=&\frac{1}{18}+\frac{b(\alpha)}{3\zeta_0},
\end{eqnarray}
where
\begin{equation}
b(\alpha)=-\frac{\pi^{\frac{d-1}{2}}}{\Gamma\left(\frac{d+5}{2}\right)d^2}
\int d\mathbf{c}_1\int d\mathbf{c}_2 \chi_H(c_1)\chi_H(c_2)
\vartheta(\mathbf{c}_1,\mathbf{c}_2),
\end{equation}
with
\begin{eqnarray}
\vartheta(\mathbf{c}_1,\mathbf{c}_2)
&=&\frac{(1-\alpha^2)(d+1+2\alpha^2)}{16}c_{12}^5\nonumber\\
&+&\frac{(d+5)-\alpha^2(d+1)+4\alpha}{4}c_{12}^3C^2\nonumber\\
&-&\frac{1+\alpha}{2}(2d+3-3\alpha)c_{12}(\mathbf{C}\cdot\mathbf{c}_{12})^2,
\end{eqnarray}
and
$\mathbf{C}=(\mathbf{c}_1+\mathbf{c}_2)/2.$ 
 The coefficient $b(\alpha)$ can be evaluated using the expression of 
$\chi_H({c})$ in the first Sonine approximation, Eq. (\ref{ec:sonine}), 
which yields
\begin{equation}
b(\alpha)=
\frac{(1 + d)(3 + d)(2 d (a_2 + 16 (-1 + \alpha) +
           15 a_2 \alpha) + 16 (-1 + \alpha) (-1 + 2 \alpha^2) +
        a_2 (7 +
           3 \alpha (-13 + 10 (-1 + \alpha) \alpha)))\Gamma[(1 + d)/2]}
{2^{1/2}\pi^{d+1/2}128 d^2 (-2 + (5 + d)/2) (-1 + (5 + d)/2) \Gamma[d/
       2] \Gamma[-2 + (5 + d)/2]}
          (1 + \alpha). 
\end{equation}
If we take into account the explicit form of $\zeta_0$ and $a_2$, given in
Eqs. (\ref{a2}) and (\ref{zeta0}) respectively, we obtain after some algebra
\begin{equation}
a_{33}=\frac{-15+7d+14d^2-3(-9+d(9+2d))\alpha+30(1+d)\alpha^2-6(9+d)\alpha^3)}
{9d(-19+2d(-7+3\alpha)+3\alpha(9+2(-1+\alpha)\alpha))}.
\end{equation}


\begin{thebibliography}{10}



\bibitem{bte05}
A. Barrat, E. Trizac, and M.~H. Ernst, J. Phys.: Condens. Matter {\bf 17},
  S2429  (2005).



\bibitem{gs95}
A. Goldshtein, and M. Shapiro, J. Fluid. Mech. {\bf 282}, 75 (1995). 



\bibitem{vne98}
T.~P.~C. van Noije, and M.~H. Ernst, Granular Matter {\bf 1}, 57 (1998).



\bibitem{wp02}
R.D. Wildman and D.J. Parker, Phys. Rev. Lett. {\bf 88}, 064301 (2002).



\bibitem{fm02}
K. Feitosa and N. Menon, 
Phys. Rev. Lett. {\bf 88}, 198301 (2002).
  

\bibitem{bt02}
A. Barrat and E. Trizac,
Granular Matter  {\bf 4}, 57 (2002).



\bibitem{sd06}
A. Santos, and J. W. Dufty, Phys. Rev Lett. {\bf 97}, 058001 (2006).



\bibitem{sn96}
H. J. Schlichtting, and V. Nordmeier, MNU {\bf 49}, 323 (1996). 



\bibitem{bmgr02}
J.~J. Brey, F. Moreno, R. Garc\'ia-Rojo, and M. J. Ruiz-Montero, 
Phys. Rev. E {\bf 65}, 011305 (2002).



\bibitem{bt03}
A. Barrat and   E. Trizac,
Molecular Physics {\bf 101}, 1713 (2003).



\bibitem{nebo97}
T.~P.~C. van Noije, M.~H. Ernst, R. Brito, and J.~A.~G. Orza, Phys. Rev. Lett.
  {\bf 79},  411  (1997).



\bibitem{bmr98}
J.~J. Brey, F. Moreno, and M.~J. Ruiz-Montero, Phys. Fluids {\bf 10},  2965
  (1998).



\bibitem{bgmr05}
J.~J. Brey, M.~I. {Garc\'{\i}a de Soria}, P. Maynar, and M.~J. Ruiz-Montero,
  Phys. Rev. Lett. {\bf 94},  098001  (2005).



\bibitem{bdgm06}
J.~J. Brey, A. Dom\'inguez, M.~I. {Garc\'{\i}a de Soria}, and P. Maynar, Phys.
  Rev. Lett. {\bf 96},  158002  (2006).



\bibitem{h83}
P. K. Haff, J. Fluid. Mech. {\bf 134}, 401 (1983). 



\bibitem{pdb03}
B. Painter, M. Dutt, and R. Behringer, Physica D {\bf 175}, 43 (2003). 

\bibitem{mima08}
C. C. Maa\ss, N. Isert, G. Maret, and C. M. Aegerter, Phys. Rev. Lett. {\bf
  100}, 248001 (2008)  


\bibitem{brm2000}
J.~J. Brey, M.~J. Ruiz-Montero, and F. Moreno, Phys. Rev. E {\bf 62}, 
5339 (2000). 



\bibitem{wm96}
D.~R.~M. Williams, and F.~C. MacKintosh, Phys. Rev. E {\bf 54}, R9 (1996).



\bibitem{plmv99}
A. Puglisi, V. Loreto, U.~M.~B. Marconi, and A. Vulpiani, 
Phys. Rev E {\bf 59}, 5582 (1999).



\bibitem{vnetp99}
T.~P.~C. van Noije, M.~H. Ernst, E. Trizac, and I. Pagonabarraga, 
Phys. Rev. E {\bf 59}, 4326 (1999).



\bibitem{ptvne01}
I. Pagonabarraga, E. Trizac, T.~P.~C. van Noije, and M.~H. Ernst,
Phys. Rev. E {\bf 65}, 011303 (2002).



\bibitem{ms00}
J.~M. Montanero, and A. Santos, Granular Matter {\bf 2}, 53 (2000).



\bibitem{mss01}
S.~J. Moon, M.~D. Shattuck, and J.~B. Swift, 
Phys. Rev E {\bf 64}, 031303 (2001).



\bibitem{gm02}
V. Garz{\'o}, and J.~M. Montanero, Physica A {\bf 313}, 336 (2002).



\bibitem{pvb05}
A. Puglisi, P. Visco, A. Barrat, E. Trizac, F. van Wijland,
Phys. Rev. Lett. {\bf 95}, 110202 (2005).



\bibitem{etb06}
M.H. Ernst, E. Trizac and A. Barrat,
J. Stat. Phys. {\bf 124}, 549 (2006).



\bibitem{peu02}
A. Prevost, D.A. Egolf, J.S. Urbach,
Phys. Rev. Lett. {\bf 89}, 084301 (2002).



\bibitem{vpbvwt06}
P. Visco, A. Puglisi, A. Barrat, F. van~Wijland, and E. Trizac, 
European Physical Journal B {\bf 51}, 377 (2006).



\bibitem{affm}
S. Auma\^\i tre, J. Farago, S. Fauve, S. Mc Namara,
European Physical Journal B {\bf 42}, 255 (2004).



\bibitem{bgmr04}
J.~J. Brey, M.~I. {Garc\'ia de Soria}, P. Maynar, and M.~J. Ruiz-Montero,
Phys. Rev. E {\bf 70}, 011302 (2004).



\bibitem{ec81}
M. H. Ernst, and E. G. D. Cohen, J. Stat. Phys. {\bf 25}, 153 (1981).



\bibitem{vk92}
N. G. van Kampen, \emph{Stochastic Proccesses in Physics and Chemistry},
(North-Holland, Amsterdam, 1992). 



\bibitem{mgsbt08}
P. Maynar, M.~I. {Garc{\'i}a de Soria}, G. Schehr, A. Barrat, and E. Trizac,
Phys. Rev. E {\bf 77}, 051127 (2008).



\bibitem{LL}
L. Landau and E. Lifshitz, {\it Physical Kinetics}, Pergamon Press, 1981.



\bibitem{cdpt03}
F. Coppex, M. Droz, J. Piasecki and E. Trizac,
Physica A {\bf 329}, 114 (2003).



\bibitem{bdr03}
J.~J. Brey, J.~W. Dufty, and M.~J. Ruiz-Montero, in 
{\it Granular Gas Dynamics}, edited by T. Poeschel and N. Brilliantov 
(Springer, Berlin, 2003).



\bibitem{gmsbt08}
M.~I. {Garc{\'i}a de Soria}, P. Maynar, G. Schehr, A. Barrat, and E. Trizac, 
Phys. Rev. E {\bf 77}, 051128 (2008).


\bibitem{02}
A. Barrat,   T. Biben,   Z. Racz,   E. Trizac,   F. van Wijland, 
J.  Phys. A: Math. Gen. {\bf 35}, 463 (2002) ; 
 A. Barrat,  E. Trizac,   M.H. Ernst,  
J. Phys. A : Math. Gen. {\bf 40}, 4057 (2007).




\end{thebibliography}
\end{document}